\documentclass[11pt]{article}

\usepackage[style=authoryear]{biblatex}
\usepackage{hyperref}
\usepackage{xcolor}
\usepackage{soul}

\usepackage{graphicx} 
\usepackage[T1]{fontenc}
\usepackage{setspace}
\usepackage[letterpaper, portrait, margin=1in]{geometry} 

\title{\LARGE{\textbf{\underline{REV}ersed \underline{I}ndexes $\approx$ \underline{VAL}ues in Wavelet Trees}} \\ \Large{\textbf{REVIVAL in Wavelet Trees}}}

\author{Xiangjun Peng\\
The Chinese University of Hong Kong\\
}
\date{Updated at 2024-2-26}

\newcommand{\revivalshort}[1]{REVIVAL}
\newcommand{\revivalfull}[1]{Reversed Indexes $\approx$ Values}
\newcommand{\revivalequal}[1]{Reversed Indexes $=$ Values}

\begin{document}

\maketitle

\begin{abstract}
    \noindent
    Data Compression (or Source Encoding) constructs indexes, by encoding information using fewer bits (than the original representation for values), to reduce the size of data storage. Therefore, if any computation needs to be performed, the compressed data needs to be uncompressed first. Since a limit of lossless data compression is expected to exist (lower-bounded by~\cite{BellTechn48/Shannon}), it is expected to expand the functionalities of compressed data under lossless compression. To this end, Succinct Data Structures are proposed~(\cite{Thesis88/Guy}) and explored, which enable queries directly on compression.
    \\
    
    \noindent
    This work presents a discovery to advance the above wisdom in a particular Succinct Data Structure: Wavelet Tree (\cite{SODA03/WaveletTree}). The discovery is first made by showing the feasibility of \revivalequal{}: for integers within $[0,2^{N})$, there exists a Wavelet Tree that its compressed indexes can be equivalent to the Leibniz Binary system~(\cite{leibniz-binary}), with only the bit reversal. Then we show how to strengthen the discovery by generalizing it into \revivalfull{}, by applying a longest common subsequence in bits and its patterns. Finally, we conjuncture potential implications of the above ideas by discussing its benefits, and modifications to the RAM model.\\
    
    \noindent
    The discovery reveals that: (1) the usability of Succinct Data Structure can be significantly expanded, by enabling Computation Directly on Compression; and (2) near-optimal lossless compression can still yield close connections with the Leibniz Binary System~(\cite{leibniz-binary}), which breeds polymorphic functionalities within a single piece of the information. This work also provides an initial analysis of the benefits from the method (and potentially other extensions), and suggests potential modifications. We conjecture that: with \underline{REV}ersed \underline{I}ndexes $\approx$ \underline{VAL}ues, everything old can be new now. \\
    
\end{abstract}

\newpage
\tableofcontents
\newpage

\begin{singlespace}

\section{Introduction}

\vspace{-4pt}

Data Compression constructs indexes to reduce the elementary size of original values, and therefore the storage cost is saved. Lower-bounded by Shannon (\cite{BellTechn48/Shannon}), a fundamental limit to lossless data compression exists, which is known as the entropy rate. For decades, an extensive amount of efforts are paid on the potential breakthrough of this limit.

This work classifies prior work into two types. One is to improve the efficiency of lossless compression; and the other is to expand the functionalities of lossless compression. The latter is the focus in this work. Succinct Data Structures, originally formalized by Jacobson (\cite{Thesis88/Guy}), are proposed to expand the functionalities of lossless compression. These data structures use an amount of space that is "close" to the entropy rate, while still retaining the functionality for efficient query operations. Therefore, queries on data can be directly performed on compression. 

However, to perform any computation, it is still needed for decompression first in Succinct Data Structures. This work centers the focus on a particular Succinct Data Structure - Wavelet Tree (\cite{SODA03/WaveletTree}). Wavelet Tree is used to store strings in compressed space. The definition of Wavelet Tree is achieved by recursively partitioning the alphabet (from the string) into pairs of subsets; the leaves correspond to individual symbols of the alphabet, and at each node a bitvector stores whether a symbol of the string belongs to one subset or the other.

This work first breaks the assumption of Wavelet Tree (or all lossless compression methods), by directly taking Leibniz Binary System (\cite{leibniz-binary}) as the reference point (rather than character encoding such as ASCII codes). We find that: the encoding patterns of Leibniz Binary System (\cite{leibniz-binary}) can be highly correlated with the formalization of Wavelet Tree (\cite{SODA03/WaveletTree}), with only the bit reversal. Therefore we derive the discovery of \revivalequal{}, which is described in Section~\ref{sec:result-revival-equal}.

This work then rolls back to the formalization of Wavelet Tree (or all lossless compression methods), by refining the reference point back to the character encoding. We find that: the above approach can be generalized to character encoding, by (1) simply accounting for common subsequence(s) in bits; and (2) utilizing these subsequences as patterns, to recover indexes to values. We also discuss extensions to other scenarios (e.g., other data types). It is described in Section~\ref{sec:result-revival-full}.

This work finally conjectures potential implications of the above ideas, by analyzing the benefits (and hypothesizing potential modifications to RAM model) for ``Indexes $\approx$ Values" principle. This work considers two viewpoints by leveraging the classification from (\cite{JDA14/WT-for-all}). We first view Wavelet Tree as a compression method, and the benefits can be directly derived. Then we view Wavelet Tree as a data structure, and the design space is discussed and we assume it deserves further investigations. It is described in Section~\ref{sec:ram-revival}.

In summary, the discovery of \revivalfull{} (\revivalshort{}) showcases the feasibility to bridge near-optimal lossless compression with the Leibniz Binary System, which makes the following three major contributions.

\vspace{-4pt}
\begin{itemize}
    \item The bridge between near-optimal lossless compression and Leibniz Binary System enables Computation Directly on Compression.
    \vspace{-4pt}
    \item The discovery expands the usability of Succinct Data Structures, and this delivers polymorphic functionalities within a single piece of the information.
    \vspace{-4pt}
    \item The bridge motivates a revamp of RAM model to support the above idea, and demonstrates potential merits of fine-grained RAM operations. 
\end{itemize}
\vspace{-8pt}

\begin{quote}
    \textit{When is a revival needed? When carelessness and unconcern keep the people asleep.}\\
    \vspace{-16pt}
    \flushright \textit{- Billy Sunday}
\end{quote}

We first give an overview of the results in this work. Then we summarize the techniques developed in this work.

\subsection{Our Results}

\subsubsection{\revivalequal{} in Wavelet Trees}
\label{sec:result-revival-equal}
\noindent
The discovery is initially made as \revivalequal{}, by extending the usage of Wavelet Tree from strings to integers. By doing so, an accidental connection is observed that: for integers within [0,2$^{N}$), there exists a Wavelet Tree that its compressed indexes can be equivalent in the Leibiniz Binary system~(\cite{leibniz-binary}), with only the bit reversal.

\subsubsection{\revivalfull{} in Wavelet Trees}
\label{sec:result-revival-full}

\noindent
The discovery is then expanded as \revivalfull{} for other types of encoding, by applying (a) common subsequence(s) in bits. Such a supplementary method breeds two opportunities. First, it can allow a more flexible mapping for the range of the to-be-compressed data. This is done by (1) extracting (a) common subsequence(s) in bits as (a) pattern(s); and (2) applying \revivalequal{} in the rest of bits. Second, it can partially break the requirement of the consecutive range for the to-be-compressed data, since the shared common subsequence(s) can have positional variations.

\subsubsection{Implications from ``Indexes $\approx$ Values" Principle}

\label{sec:ram-revival}
\noindent
Based on the above ideas, this work discusses potential benefits, and suggests a potential revamp of RAM model. Our discussion takes WT as a motivating example, and classifies our analysis into two cases. We first view WT as a compression method, and the benefits can be derived directly. Then we view WT as a data structure, and suggest potential modifications of RAM model: so that we can support both queries and computation directly on compression.

\subsection{Technique Outline}

    \noindent
    Though the discovery of \revivalshort{} is empirical, there are the following techniques can be derived for further investigations of other instances in the ``Indexes $\approx$ Values" principle.\\

    \noindent
    \textbf{$\bullet$ Input-bounded Range of Available Values}: the first step for the usage of \revivalshort{} (or other techniques) requires the bounded range of all available values, as the preliminary knowledge. This is used to determine how \revivalshort{} shall play a role in this bounded range, and furthermore how the longest common subsequence in bits shall be extracted (if needed).\\

    \noindent
    \textbf{$\bullet$ Conditional Partitions of the Alphabet}: the second step for the usage of \revivalshort{} (or other techniques) requires a set of conditions to be determined, which are used for encoding all values. The key to decide these conditions shall be closely related to how these values are encoded, and they are expected to be closely correlated.\\

    \noindent
    \textbf{$\bullet$ Executions on ``Indexes $\approx$ Values" Principle}: the final step for the usage of \revivalshort{} (or other techniques) requires the abstraction of the above information, for the fine-grained operations over the compressed sequence of bits. This can vary by leveraging different viewpoints, and how different functionalities shall be supproted.

\section{Background}

Wavelet Tree (WT) is a Succinct Data Structure, that can support $rank$ and $select$ operations efficiently, to store strings in compressed space (\cite{SODA03/WaveletTree}).

\vspace{-4pt}

\subsection{Wavelet Tree Definition}

\begin{figure}[!h]
    \centering
    \includegraphics[width=\linewidth]{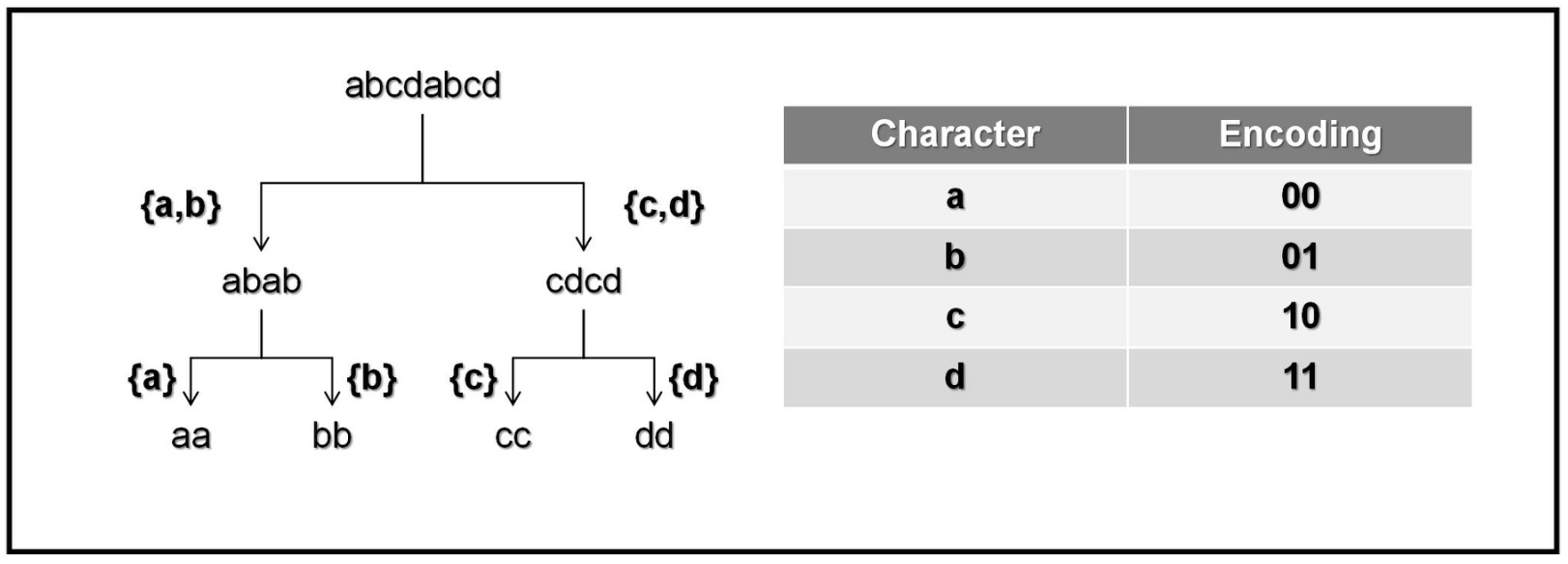}
    \vspace{-4pt}
    \caption{An example Wavelet Tree of the string ``abcdabcd".}
    \label{fig:wt-example}
\end{figure}

A WT is a data structure that recursively partitions a stream of characters into two parts, until homogeneous data are left. The encoding scheme is dependent to the partition of alphabet (and its subsets). Figure~\ref{fig:wt-example} gives out an example of WT from the string ``abcdabcd".

\vspace{-4pt}

\vspace{-4pt}

\subsection{Bitmap from Wavelet Tree}

\begin{figure}[!h]
    \centering
    \includegraphics[width=\linewidth]{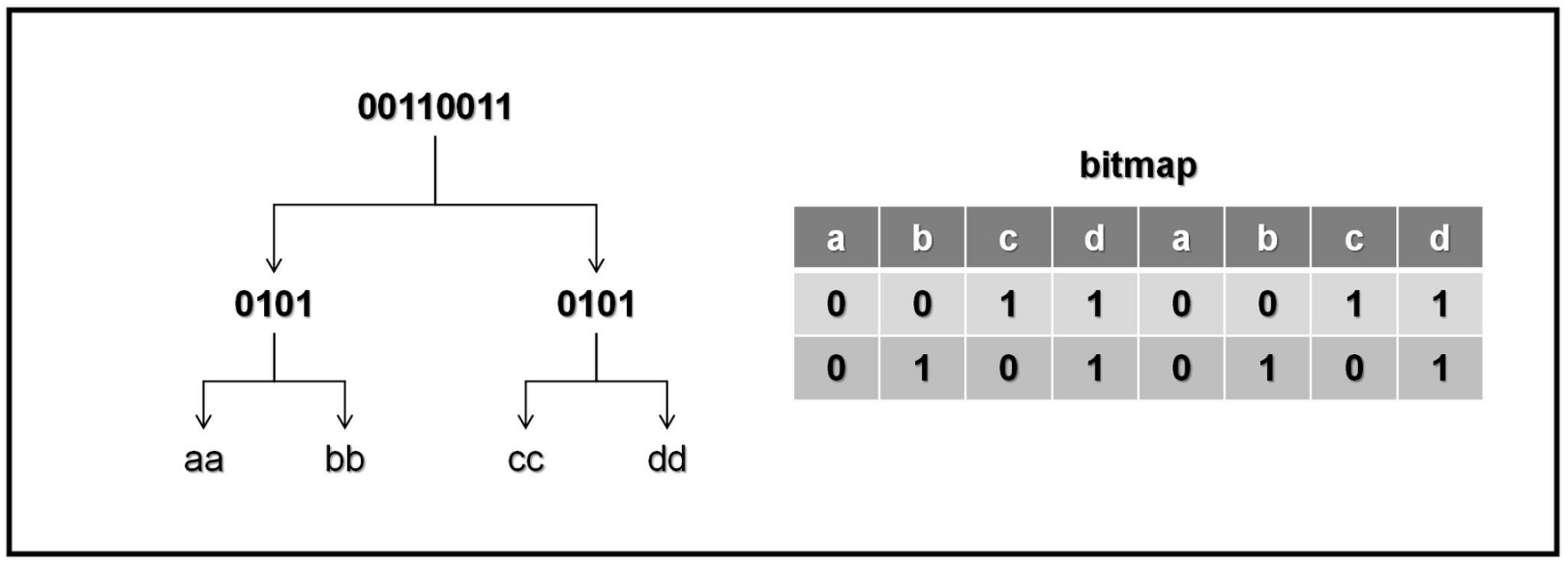}
    \vspace{-4pt}
    \caption{The corresponding bitmap from Figure~\ref{fig:wt-example}.}
    \label{fig:wt-bitmap}
\end{figure}

A bitmap from WT is to deliver the encoding results level by level, and every index can be viewed vertically (from top to down). Figure~\ref{fig:wt-bitmap} gives out the demonstrated bitmap from Figure~\ref{fig:wt-example}.

\section{\revivalequal{}}

We first introduce \revivalequal{}. We make the discovery by changing the reference point from character encoding into Leibniz Binary System (\cite{leibniz-binary}).

\subsection{Leibniz Binary System Made WT}

\begin{figure}[!h]
    \centering
    \includegraphics[width=\linewidth]{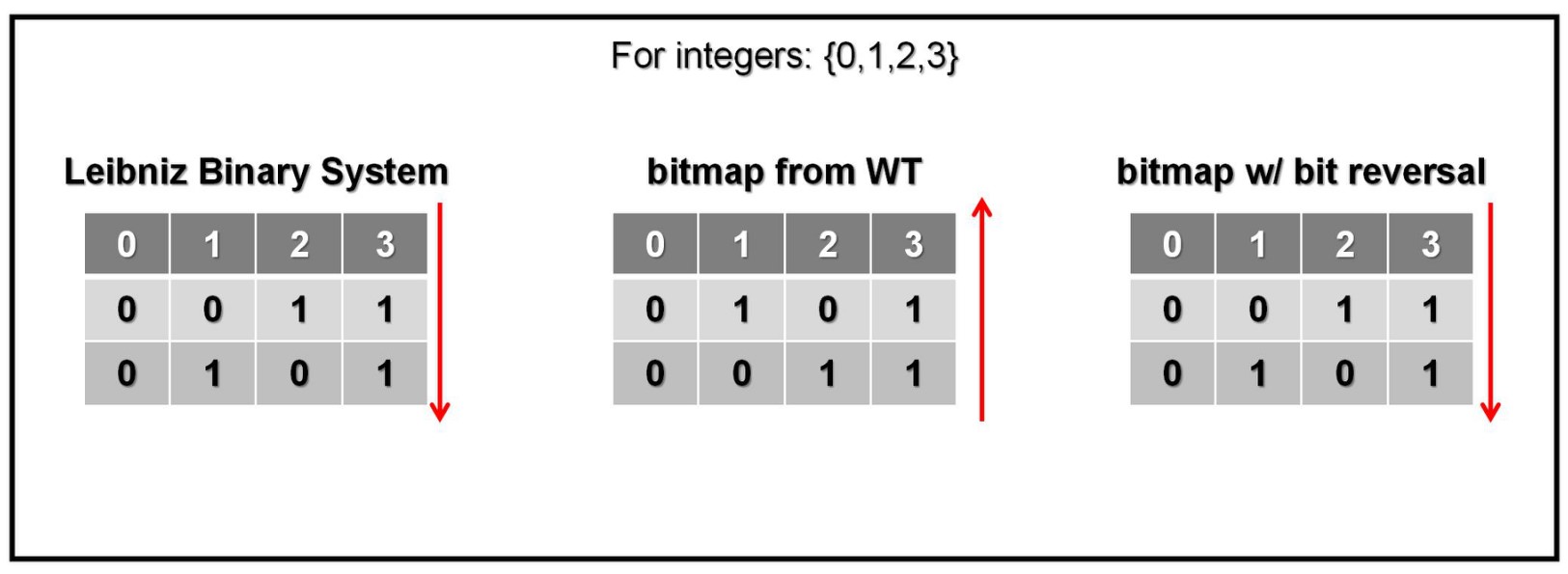}
    \vspace{-4pt}
    \caption{An example of \revivalequal{} using integers within $[0, 2^{2})$.}
    \label{fig:revival-equal}
\end{figure}

The "\revivalequal{}" is described hereby: for integers within $[0,2^{N})$, there exists a Wavelet Tree that its compressed indexes can be equivalent to the Leibniz Binary system~(\cite{leibniz-binary}), with only the bit reversal. Figure~\ref{fig:revival-equal} gives an example to demonstrate the idea.

\subsection{Generalization to integers within $[0,2^{N})$}

The above discovery can be easily generalized. This is because the encoding scheme from Leibniz Binary System (\cite{leibniz-binary}) is a natural fit with the definition of WT. To obtain the above results in a generalized form (i.e. for integers within $[0,2^{N})$), it is straightforward to have the following method derived:

\begin{quote}
\textbf{Sort the alphabet, and then perform the partition for WT by putting smaller ones into a subset, and the rest into the other subset.}    
\end{quote}

There are a few notes regarding the above generalization of \revivalequal{}.

First, note that this may not be the only method to derive parts of the same results, but the above one is considered as the most general one. This is because the above rule is derived based on binary carry in Leibniz Binary System. More specialized forms are expected to be delivered also.

Second, the lower bound of the input data range has to remain as zero so that the proposed method can function. This is expected since WT encoding requires one of all elements from the alphabet to encoded as ``0"s, and therefore enforces the inclusion of zero.

Third, similar results may be derived if we change the overall range of the input data. However, the regulations to partition the alphabet can not be succinct enough, and the derived results can only be an approximation of \revivalequal{}. More instances, from slight changes of the input range, can be derived. We leave this part to the next section.

\section{\revivalfull{}}

With the strict definition of \revivalequal{}, we present a more general form called \revivalfull{}.

\subsection{Common Subsequence in Bits Made \revivalfull{}}

\begin{figure}[!h]
    \centering
    \includegraphics[width=\linewidth]{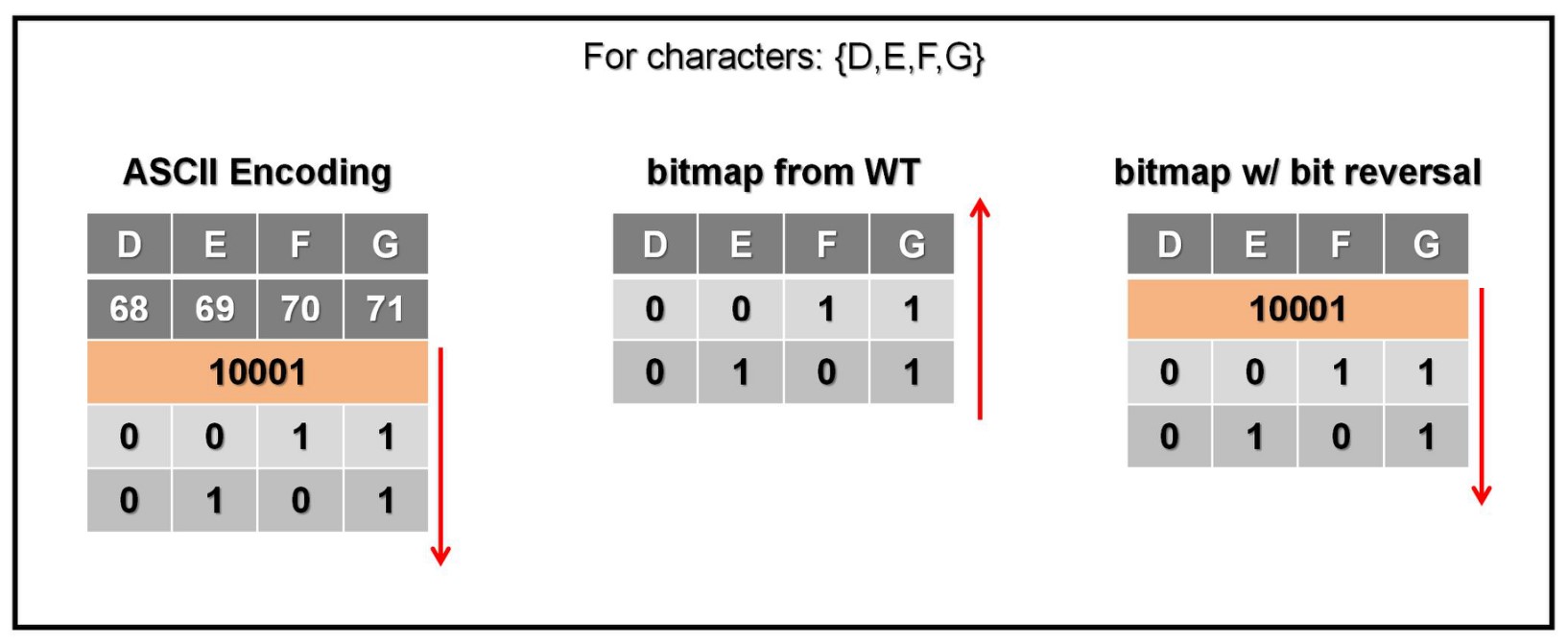}
    \vspace{-4pt}
    \caption{An example of \revivalfull{}{} using characters within $[D, G]$ in ASCII encoding. The highlighted ``10001" is the shared common subsequence of all characters in bits.}
    \label{fig:revival-full}
\end{figure}

The "\revivalfull{}" is described hereby: when leveraging parts of the bit sequence via \revivalequal{}, one or (several patterns) patterns can be used to connect bitmap from WT with the exact values in different encoding schemes. Figure~\ref{fig:revival-full} gives an example to demonstrate the idea in ASCII encoding, and we elaborate more on this example.

For $[D, G]$ in ASCII encoding, all values share the common bit subsequence ``10001" in binary. Therefore, after following the principle of \revivalequal{}, we can obtain the bitmap and the reversed bits can be equalized with parts of the exact values. To fully recover the values, the only job is to supplement the common subsequence (i.e. ``10001").

\subsection{Extensions in \revivalfull{}}

It is expected that \revivalfull{} can be extended in a variety of aspects.

First, note that the usage of common subsequences can be generalized, since there can be several common subsequences within the input data range. Hence, the number of these subsequences determine the number of bit patterns, and these patterns are used to isolate so that the rest of bits can be used via \revivalequal{}.

Second, though the demonstrated example only covers character encoding in ASCII encoding, it is expected to be generalized to other types of value encoding. Particularly, for floating-point numbers, we assume dyadic scaling is more suitable one for the discovery.

Third, the discussion so far does not cover the impacts of the sign system, but the inclusion of a sign system is expected not to impact the correctness of our method for \revivalequal{}, as long as the value of zero is included.

\section{Implications from ``Indexes $\approx$ Values" Principle}

\vspace{-4pt}

We discuss benefits from (and potential modifications to) RAM, so that we can support the usage of \revivalfull{} (and their potential variants). The key to support the polymorphic functionalities is to enable the retrieval of these values efficiently. We leverage \revivalfull{} as an example: based on the usage of WT, we analyze (and provide potential modifications) using RAM for \revivalfull{}. We first view WT as a compression method; and then we view WT as a data structure.

\vspace{-4pt}

\subsection{Viewpoint as a Compression Method}

When taking WT as a compression method, the benefits are straightforward and two-fold. First, every index is directly packed together, which reduces the overhead in terms of data transfer. Second, the processor only requires bit manipulation to conclude the de-compression (in \revivalfull{}, the bit reversal or constant values for recovering the values based on bit patterns): it saves costs in common de-compression (e.g., lookup costs).
\vspace{-4pt}

\subsection{Viewpoint as a Data Structure}

Taking WT as a data structure requires decent modifications to RAM for the usage of \revivalfull{}, and we conjecture that there are three parts. First, it requires dual-address modes for indexes and values respectively. Second, it requires level-oriented gather supports by using levels of WT as breakpoints, so that values can be retrieved. Third, though bit manipulation (and constant values for recovering the values based on bit patterns) can be performed within the processor, there can be benefits by integrating these functionalities in some cases (e.g., Processing-In-Memory Paradigm). We leave this to the future investigations.

\vspace{-4pt}

\section{Conclusions}

\vspace{-4pt}

This work describes a discovery to bridge near-optimal lossless compression with Leibniz Binary System. It (1) makes Computation Directly on Compression feasible; and (2) enables polymorphic functionalities (i.e., efficient queries and computation) within a single piece of the information. This work also provides an initial analysis of the benefits from the method (and potentially other extensions), and suggests potential modifications. We conjecture that: with \revivalfull{}, everything old can be new now. 

\vspace{-4pt}

\end{singlespace}

\printbibliography

\appendix

\section*{Review Outcome from ACM STOC 2024}

\begin{itemize}
    \item This preprint was submitted to ACM STOC 2024, and rejected for the following reason:
    \begin{itemize}
        \item \textit{This paper discusses Wavelet Trees and a manner for splitting characters in them based on binary notation. However, no new theorems are presented. Thus, the PC concluded the paper is not appropriate for the CS theory conference STOC.}
    \end{itemize}
    
\end{itemize}

\section*{Personal Comments}

\begin{itemize} 
    \item Hilarious. :-) 
\end{itemize}

\begin{figure}[!h]
    \centering
    \includegraphics[width=0.7\linewidth]{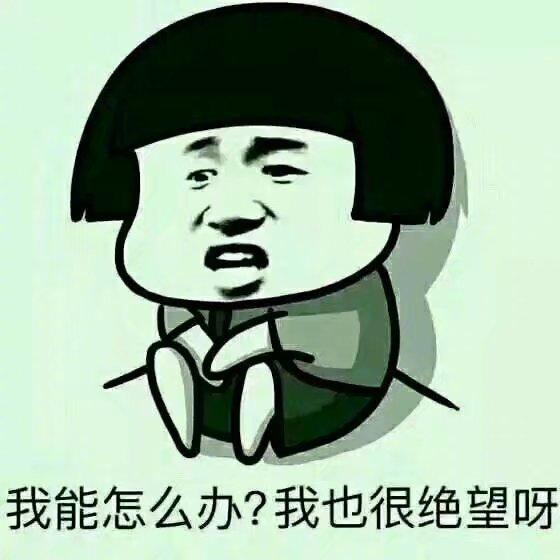}
\end{figure}

\end{document}